\documentclass{jaa}
\usepackage{natbib}
\usepackage{aas_macros}

\usepackage[utf8]{inputenc}
\usepackage[T1]{fontenc}
\usepackage{makecell}

\usepackage{graphicx} 
\usepackage{amsmath,amssymb}

\usepackage{xspace}	
\usepackage{lineno}
\usepackage{natbib}
\usepackage{xcolor}
\usepackage[colorlinks=true,citecolor=blue]{hyperref}

\newcommand{\sw}[1]{\texttt{#1}}
\newcommand{\sqd}{\ensuremath{\mathrm{deg}^2}}

\newcommand{\strain}{\{x(t)\}}

\newcommand{\stratmass}{\ensuremath{S_\mathrm{M}}}
\newcommand{\stratprob}{\ensuremath{S_\mathrm{2DP}}}

\newcommand{\stratmpthree}{\ensuremath{S_\mathrm{3DPM}}}
\newcommand{\stratmassfill}{\ensuremath{S_\mathrm{3DMF}}}

\begin{document}\sloppy

\title{Efficacy of Galaxy Catalogues for following up gravitational wave events}

\author{Tamojeet Roychowdhury\textsuperscript{1*}, Harsh Choudhary\textsuperscript{2,3}, Varun Bhalerao\textsuperscript{3}, David O. Cook\textsuperscript{4}, Viraj Karambelkar\textsuperscript{4}, Mansi M. Kasliwal\textsuperscript{4}, Harsh Kumar\textsuperscript{3,5}, Surhud More\textsuperscript{6} and Gaurav Waratkar\textsuperscript{3} }
\affilOne{\textsuperscript{1}Department of Electrical Engineering, Indian Institute of Technology, Bombay, Mumbai 400076, India\\}
\affilTwo{\textsuperscript{2}Department of Computer Science, Czech Technical University, Prague 12000, Czechia\\}
\affilThree{\textsuperscript{3}Department of Physics, Indian Institute of Technology, Bombay, Mumbai 400076, India\\}
\affilFour{\textsuperscript{4}California Institute of Technology, 1200 E. California Boulevard, Pasadena, CA 91125, USA\\}
\affilFive{\textsuperscript{5}Center for Astrophysics | Harvard \& Smithsonian, 60 Garden Street, Cambridge, MA 02138-1516, USA\\}
\affilSix{\textsuperscript{6}Inter-University Center for Astronomy and Astrophysics, Pune 411007, India}

\twocolumn[{

\maketitle

\corres{tamojeetroychowdhury@gmail.com}


\begin{abstract}
The detection of gravitational waves (GW) by the LIGO-Virgo-KAGRA (LVK) network has opened up a new era in astrophysics. The identification of the electromagnetic counterparts of GW sources is crucial for multi-messenger astronomy, one way of which is to use galaxy catalogues to guide optical follow-up observations. In this paper, we test the utility of galaxy-targeted approach with mass prioritised galaxy ranking for the ongoing LIGO O4 run. We have used the simulated results for the expected LIGO O4 events and the NED-LVS galaxy catalogue and based our study for small field of view telescopes, specifically the GROWTH-India Telescope (GIT). With the increase in sensitivity of LIGO/Virgo in the ongoing observing run O4, the expected number of total detections have gone up but most of these are also now poorly localised. We show that a larger volume covered in the same field-of-view (FoV) on the sky results in a large increase in the total number of galaxies in each FoV. A significant top-heaviness is observed in the mass-ranked list of galaxies, which still number to a few thousand in most cases. At larger distances, such high numbers of deep follow-up observations are infeasible in most cases rendering galaxy catalogues useful in limited cases, but these are still useful at lower distances where LVK detectors are currently sensitive and where galaxy completeness is higher. We also explore the effect of mass-filling to account for galaxy catalogue incompleteness at large distances. If mass-filled probabilities are considered as the metric for ranking and coverage, we find that the conventional 2D probability search performs better than a 3D galaxy catalogue (without mass-filling) based search at distances larger than 300 Mpc (upto which NED-LVS is $\sim 70\%$ complete), and using 3D mass times probability in each tile performs better for nearby events.
\end{abstract}


}]

\doinum{12.3456/s78910-011-012-3}
\artcitid{}
\volnum{000}
\year{0000}
\pgrange{13}
\setcounter{page}{1}
\lp{13}

\section{Introduction} 
\label{sec:intro}

Advanced gravitational wave (GW) detectors including Laser Interferometer Gravitational-Wave Observatory \citep[LIGO;][]{Aasi15}, Virgo \citep{Acernese15} and Kamioka Gravitational Wave Detector \citep[KAGRA;][]{Aso13} have revolutionised multi-messenger astronomy by enabling the direct observation of cosmic events like neutron star mergers, thereby offering a new way to study the universe through the detection of GWs alongside traditional electromagnetic (EM) signals. 
 
The detection of gravitational waves from a binary neutron star (BNS) merger GW170817 \citep{2017PhRvL.119p1101A} accompanied by an electromagnetic counterpart seen across the electromagnetic spectrum \citep{2017ApJ...848L..13A,2017ApJ...848L..12A,coulter2017,2017ApJ...848L..15S,2017Natur.551...71T} has provided unprecedented insights into the nature of extreme astrophysical phenomena \citep{2017ApJ...850L..19M, 2017Natur.551...67P, 2017Natur.551...80K, 2018PhRvL.120q2703A,kns+17}.

However, subsequent searches for electromagnetic counterparts to BNS or Neutron Star -- Black Hole (NSBH) mergers failed to find any counterparts \citep{ztfo4,2020ApJ...905..145K,2020MNRAS.497.5518A,2021ApJ...912..128P,2022ApJ...924...80U,2022MNRAS.516.4517K,2023ApJ...942...99G,2024ApJ...964..149F,2024ApJ...976..123W,2025ApJ...980..207R}. Two of the key challenges in such searches are the large localisation areas spanning hundreds to thousands of square degrees which requires a large number of telescopic images to cover the area, and the larger sources distances which require longer observations at each pointing to detect fainter sources \citep{Weizmann_2023,ztfo4}.

The distribution of short Gamma Ray Bursts (SGRBs) arising from merging neutron stars follows the host galaxy masses \citep{berger2014,2022ApJ...940...56F}. As one can expect a similar distribution for BNS events, a basic follow-up strategy would preferentially point telescopes at galaxies included in the localisation volume, avoiding pointing to the relatively empty space in between \citep[see for instance][]{Ducoin_2020}. Also, \cite{Nugent_2022} also found that a majority of SGRB hosts are star-forming.
Indeed, simulations support this idea and show a strong correlation between the galaxy mass and the probability of detecting a BNS or NSBH merger in that host \citep{Mapelli2018,Artale2019,artale2020mass}. These simulations show as well that the merger rates are correlated to star formation rate (SFR).

The possibility of using galaxy catalogues for GW follow-up has been discussed in the literature for over a decade \citep[see for instance][]{Nicholl_Andreoni_2024}. \cite{nkg12} and \cite{2014ApJ...789L...5K} simulated the effect of using such catalogues in follow-up observations and found that using galaxy catalogues can decrease the EM search regions by two to three orders of magnitude, but noted that galaxy catalogues were grossly incomplete. \citet{Hanna_2014} concluded that catalogues with completeness as low as 33\% are still better than not using galaxy catalogues at all, while more complete catalogues can give a boost of $\sim$300\% to the probability of success. \citet{kanner2012a} discussed  a galaxy targeting strategy for follow-up observations with the \emph{Neil Gehrels Swift Observatory}. This was expanded upon in \citet{gehrels2016galaxy} where they found that imaging 10--20 galaxies in the localisation volume would cover about 50\% probability of containing the EM counterpart in GW observing runs, thereby decreasing the necessary \emph{Swift}/XRT pointings by one to two orders of magnitude. Such strategies have borne success: the discovery of SSS17a/AT2017gfo resulted from a galaxy--based search undertaken by \citet{coulter2017}.

However, the increasing sensitivity of the GW detectors has led to higher typical distances for detected events, challenging the galaxy catalogue strategy in two ways. Firstly, the galaxy catalogues have limited completeness, and do not include a large number of galaxies beyond $\sim 200$~Mpc. Secondly, the small localisation volumes in earlier observing runs such as O2 included only a handful of galaxies \citep{gehrels2016galaxy}. The new large GW localisation volumes in O4 and beyond may contain thousands of galaxies, such that imaging each of them may end up requiring the telescopes to cover most of the projected localisation region. How should the galaxy catalogue strategy be revised in this context, and is it even useful any more? In this paper, we evaluate the efficacy of the galaxy catalogue pointing strategy for current and future follow-up programs.

The paper is organised as follows. We discuss some available galaxy catalogues and select one for our simulations in \S\ref{sec:galaxycatalogue}. In \S\ref{sec:strategies} discuss a few strategies for selecting the order of observations. We consider simulated follow-up observations with the GROWTH-India Telescope and WINTER in \S\ref{sec:simobs}, and discuss the results in \S\ref{sec:results}.  We conclude with a discussion in \S\ref{sec:discuss}.

\section{Galaxy catalogues}
\label{sec:galaxycatalogue}

Implementing a galaxy targeting follow--up strategy involves multiple steps: (a) identifying galaxies within a GW localisation volume, (b) ranking the galaxies in priority order, (c) creating an observing schedule.

Step (a) is the most straightforward: given a GW localisation volume and a galaxy catalogue, one can easily identify the galaxies contained in a certain fraction of the localisation volume \citep{Singer_2016}. To ensure reasonable coverage without increasing the volume by large amounts, we look at galaxies within 99\% of the localisation volume. Step (b) requires assigning individual galaxies weighted 3D probabilities of containing the EM source. As alluded to in \S\ref{sec:intro}, one expects the probability of containing the source to be proportional to galaxy mass. 
 Thus, in order to use a galaxy catalogue for planning follow-up observations of GW sources, the catalogue needs to satisfy the following requirements: first and foremost, the catalogue should be as as complete as possible, which is currently primarily limited by the availability of accurate redshifts. Second, it should provide information about galaxy masses and star formation rates. In addition to these galaxy properties available from catalogues, one also needs to consider the position of the galaxy within the localisation volume: a lighter galaxy at a high probability point might be more valuable than a heavier galaxy in a lower probability region. Finally, step (c) requires knowledge of observing limitations, grouping of galaxies within a single exposure by the telescope, etc. which we discuss in \S\ref{sec:simobs}.

Given the potentially large gains of a galaxy--targeting strategy, many efforts have been directed towards making galaxy catalogues like Gravitational Wave Galaxy Catalog \citep[GWGC;][]{gwgc}, Census of the Local Universe \citep[CLU;][]{kasliwal2011a,cook2019}, Compact Binary Coalescence Galaxy \citep[CBCG;][]{kopparapu2008a} Galaxy List for the Advanced Detector Era \citep[GLADE;][]{glade}, GLADE+ \citep{D_lya_2022}, the Heraklion Extragalactic Catalogue  \citep[HECATE;][]{Kovlakas_2021}
and the NASA/IPAC Extragalactic Database (NED) Local Volume Sample \citep[NED-LVS;][]{cook2023completeness}.

Galaxy catalogues are limited in terms of their range and completeness. For instance, GWGC and CBCG were limited only to 100~Mpc, appropriate for the limited range of GW detectors of the time. Thanks to efforts from various groups around the world, galaxy catalogues have improved over the years. Subsequent catalogues like CLU, GLADE, and GLADE+ pushed these ranges to 200~Mpc and beyond as the range of GW detectors improved. However, the range of GW detectors has increased even further: \citet{Abbott_2020} projected the O4 median BNS mergers detection range to be ${352.8}_{-9.8}^{+10.3}$ Mpc, and median NSBH mergers range to be ${621}_{-14}^{+16}$ Mpc. Indeed, a look at GraceDB\footnote{The distances and sky localisations of all GW events can be found at \url{https://gracedb.ligo.org/superevents/public/O4/}} shows that most BNS and NSBH events in O4 had median distances above 200~Mpc, extending up to a gigaparsec. For such events, targeted follow--up strategies in must rely on galaxy catalogues that contain galaxies to much higher distances.

A recent catalogue which meets these requirements is the NED-LVS~\citep{cook2023completeness}. Compiled from the NASA/IPAC Extragalactic Database\footnote{The NASA/IPAC Extragalactic Database (NED) is funded by the National Aeronautics and Space Administration and operated by the California Institute of Technology.}, it contains $\sim$1.9~million objects to a distance of 1000~Mpc. The catalogue is $\sim$70\% complete up to a distances of 300--400~Mpc, and is 10\%--20\% more complete than GLADE and HECATE beyond 80~Mpc. Further, for galaxies that are brighter than $L_*$ (the characteristic luminosity in the galaxy luminosity distribution function), the catalogue is nearly 100\% complete to about 450--500~Mpc. Since the integrated near-infrared luminosity is proportional to the stellar mass in most galaxies (ignoring sources like active galactic nuclei), this means that for the most massive galaxies (which are most likely to contain the GW event progenitor), the catalogue is almost complete till upto 450 Mpc.

Follow--up of GW events is a key driver for the creation of this catalogue, and it forms the underlying sample for the NED gravitational-wave follow-up service (NED-GWF)\footnote{\url{https://ned.ipac.caltech.edu/NED::GWFoverview/}}. The catalogue contains all galaxies included in the vast majority of earlier catalogues, is regularly updated with new objects and includes key galaxy parameters like masses and SFR, needed for ranking the galaxies as discussed above. Hence, we use this as the base catalogue in this work.

\section{Simulated observations}\label{sec:simobs}

As mentioned in \S\ref{sec:galaxycatalogue}, actual observations involve subtleties that need to be properly accounted for when evaluating observing strategies. 
For images containing multiple galaxies, the total probability covered in each image will vary. Survey telescopes often have pre--defined grids of ``tiles'' on the sky, which may have overlaps or gaps. A galaxy falling in a gap will not be seen by that telescope at all, unless a special exception is made while scheduling. In order to account for such effects, we consider a set of simulated GW events (\S\ref{sec:gwevents}) and simulate follow-up observations with two telescopes: the Growth-India Telescope (GIT) and the Wide-Field Infrared Transient Explorer (WINTER). 

Some factors which impact follow-up observations include the visibility of the fields at the telescope site based on its location \citep{Srivastava2017}, the time of the event during the night \citep{Rana_2017}, distance of the localisation region from the Sun, etc. This complicates the simulations, but such effects have already been explored in other works. Several current algorithms use a greedy approach to scheduling, selecting the tile enclosing the highest probability at each time step. This method is used in \texttt{gwemopt} scheduler \citep{Coughlin_2019} and in \texttt{Astroplan} \citep{Morris_2018}. Some other observatories such as the Las Cumbres Observatory (LCO) and Atacama Large Millimetre Array (ALMA) use an integer linear programming based approach \citep{lampoudi2015, SOLAR201690} that makes the scheduling problem computationally easier while also accounting for scientific priorities and telescope capacity by formulating the problem with discrete integer variables, linear constraints and an objective function to be optimised. This was more recently implemented for time-domain imaging surveys by the Zwicky Transient Facility \citep{Bellm_2019}, as well as explored for GW follow-up by future missions such as UltraViolet EXplorer \citep{Singer_2025}. While these works primarily focus on scheduling and observability constraints, they typically do not utilise other information such as galaxy catalogues for GW or transient follow-ups. 

The scheduling problem is thus in general complex and any GW event in practice would have additional constraints in real-time observation that are difficult to account for in simulations. Instead, we take a simplified approach towards ``observability'' of an event. First, selection of the tile grid for a telescope automatically rejects events that whose entire 99\% localisation area is never visible (in our case, too far south). Next, we consider that a typical observing scheme with few--minute exposures giving about 100 exposures in a night. 

Thus, we simply assume that for the telescopes under consideration, observations can cover the hundred tiles with the highest weights for respective schemes from \S\ref{sec:strategies}. This simplistic assumptions allows us to focus on comparing the impact of galaxy catalogues without worrying about other constraints. If other constraints limit the number of observations possible, the total probability covered by any given strategy decreases, and typically so does the difference in outcomes from different strategies.

\subsection{GROWTH-India Telescope}\label{sec:git}
The first telescope we consider is the GROWTH-India Telescope \citep[GIT;][]{Kumar2022} --- a 0.7~m optical telescope located at the Indian Astronomical Observatory at Hanle, Ladakh, India. This robotic telescope is equipped with SDSS u$^\prime$ g$^\prime$ r$^\prime$ i$^\prime$ z$^\prime$ and has a $0.7^\circ$ diameter circular field of view (FoV), covering 0.38~\sqd. on the sky. GIT can observe targets up to an altitude of about $25^\circ$ above the horizon, corresponding to a minimum observable declination of $-33^\circ$. For GW follow--up observations, GIT uses a pre--defined grid of tiles on the sky\footnote{GIT tiling scheme: \url{https://sites.google.com/view/growthindia/about/tiling-scheme}} \citep[also see][]{2022MNRAS.516.4517K}, which we use in our simulations.

\subsection{WINTER}\label{sec:winter}
Next we select a wider field telescope: the Wide-Field Infrared Transient Explorer \citep[WINTER;][]{Frostig_2022} --- a near-infrared transient follow-up telescope designed specifically for studying binary neutron star merger kilonovae. The telescope observes in the Y, J, and short-H bands, and has a 1.2~\sqd\ FoV.  
WINTER too utilises a pre--defined all--sky grid for observations, which we use in our simulations below. 
Comparing simulated observations with this nearly double FoV as compared to GIT allows us to test how the differences between various schemes change with the FoV.

\subsection{GW event sample}\label{sec:gwevents}
In order to check the efficacy of galaxy catalogue-based follow-up, we need a sample of GW events. We have based our analysis on simulations by 
\citet{Weizmann_2023}. The work presented simulations of two LIGO/Virgo/KAGRA observing runs, O4 and O5, that are grounded in the statistics of O3 public alerts. This set includes a total of 1132 neutron star merger events (BNS or NSBH), selected using the criteria that the mass of the smaller component is less than $3 M_\odot$.

The median 90\% localisation area for O4 is predicted to be ${1820}_{-170}^{+190}$~deg$^2$ and ${1840}\pm150$~deg$^2$ for BNS and NSBH events respectively,  which is higher than what was observed in O3 \citep{Abbott_2020}. The median luminosity distance is also expected to go up to $353\pm10$~Mpc and ${621}_{-14}^{+16}$~Mpc for BNS and NSBH respectively. The injections themselves are random in space, without considering the locations of actual galaxies. However, given the large number of galaxies contained in each localisation volume (\S\ref{subsec:ngal}), this factor will not impact the results of our study.
With this increased distance to the GW events and larger sky localisations, finding an EM counterpart will become harder. This makes it more important to study and quantify the efficacy of using galaxy catalogues for this search.

Since we are applying a 100-image criteria for the two telescopes as discussed above, we expect to cover up to 38 and 120~deg$^2$ for any event respectively. Thus, we choose to ignore events that are very poorly localised, where the coverage by these telescopes will be very small. We conservatively set this cutoff at events with 99\% area under 3600~deg$^2$. We note that the results of this study are not highly sensitive to this cutoff. 

This leaves us with 329 events in the catalog, with distances ranging from 67~Mpc to 910~Mpc, and a sample median of 302~Mpc. The median 99\% area is 1596~\sqd, close to half of our localisation area cutoff of 3600~\sqd. For each of these events, we simulated observations using various observing schemes to be discussed in \S\ref{sec:strategies}, and compare the results.

\subsection{Number of Galaxies in the Localisation Volume}\label{subsec:ngal}
\begin{figure}[htbp]
    \centering
    \includegraphics[width = \columnwidth]{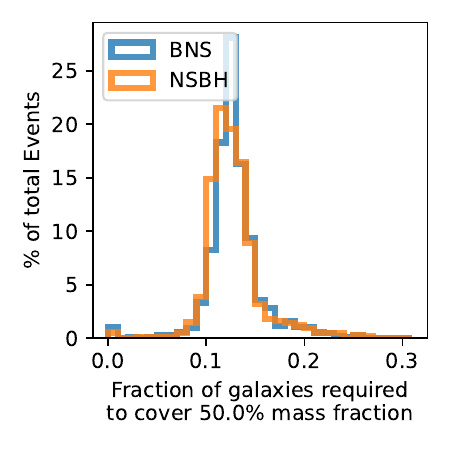}
    \includegraphics[width = \columnwidth]{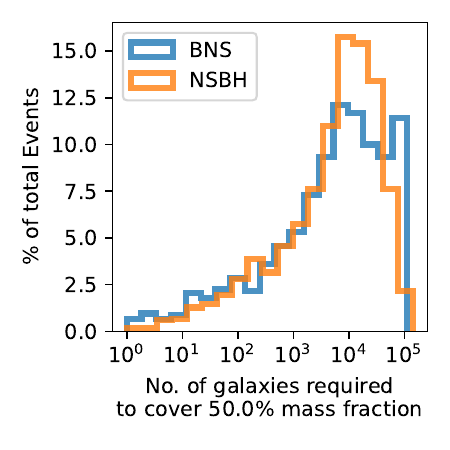}
    \caption{\textit{Top panel:} The distribution of the fraction of galaxies that need to be covered to include 50\% of the mass fraction in a mass-sorted list of galaxies, over the complete set of simulation events. Here, it is clear that for typical events, both the BNS and NSBH plots peak at around 12\% which shows that there is a significant top-heaviness in the mass-sorted galaxy list
    \textit{Bottom panel:} The distribution of the total number of galaxies needed to be covered to cover 50\% mass fraction in the mass-sorted list. We see the localisation volume contains $10^3$ to $10^5$ known galaxies for most events, in addition to any galaxies without catalogued redshifts.}
    \label{fig:top-heavieness}
\end{figure}

For each event from the above sample, we select the galaxies within the 99\% localisation volume. Exploring the mass distribution of the selected galaxies, we find that the heaviest 12--15\% galaxies account for 50\% of the mass within the localisation volume (Figure~\ref{fig:top-heavieness}, top panel). The narrow width of this feature indicates that the galaxy mass function in every volume considered is similar. In case the source is nearby or the localisation volume is small, this converges to the findings of \citet{gehrels2016galaxy} where few tens of galaxies were found to be sufficient for covering $\sim50$\% of the total light.

However, in the current and future runs of the IGWN (international gravitational wave network), the typical distances to compact binary sources are much higher, leading to larger localisation volumes. These volumes often contain tens of thousands of galaxies, and even the heaviest 12--15\% of them (needed to cover 50\% of the mass) now adds up to $10^3$--$10^5$ galaxies, as shown in the bottom panel of Figure~\ref{fig:top-heavieness}. This is in stark contrast to \citet{gehrels2016galaxy}: where due to the lower sensitivity of the GW detectors, only nearby events were detected and thus volumes were smaller. 

A subtle point to note here is that the number of galaxies to be observed does not equal the number of telescope pointings: based on the telescope field of view, an image from a single telescope pointing may contain multiple galaxies, and the number of pointings/images required to cover 50\% galaxy mass may be lower. 

\section{Exploring follow-up strategies for O4}\label{sec:strategies}

We now consider four strategies for ordering the galaxies to be observed (Table~\ref{table:comparisons}). For the rest of this work, we have assumed that the merger rates trace galaxy mass, hence have used mass (rather than SFR or other properties) as the main galaxy property to be considered in weighing galaxies for follow--up. 

Current LVK event alerts are accompanied with a localisation sky map\footnote{For details see \url{https://emfollow.docs.ligo.org/userguide/tutorial/skymaps.html}.}, which divides the sky into a number of pixels based on the \sw{healpix} pixelisation \citep{healpix} scheme and assigns to each pixel the probability of the event having occurred in that pixel. Along each pixel, there is also a probability density of that event having occurred at varying radial distances: typically expressed as the mean and standard deviation for a normal distribution along the line of sight for each pixel\footnote{For details see \url{https://emfollow.docs.ligo.org/userguide/content.html\#sky-localization}.}. The probability can thus be seen as both a 3D distribution in space (henceforth denoted $p_{\rm{3D}}$), or as a marginalised 2D distribution over the celestial sphere (henceforth, $p_{\rm{2D}}$). For our simulations, we use a healpix grid with $N_\mathrm{side} = 512$, and map each galaxy to a healpix pixel. We make a simplifying assumption that each galaxy lies within a single healpix pixel, which may be an issue for large nearby galaxies included within the localisation volume of an event. However, given the $\sim$degree fields of view used in our simulated observations, most galaxies would be included in a single tile and hence in a single simulated image, and this is a safe assumption. Alternately, when more than one galaxy lies within a single pixel, their weights would simply add up in the total weight/priority of that tile.

\subsection{2D Probability Based Pointing}\label{sec:strat_prob}
Our baseline method is to sort the tiles by the 2D probability obtained from the GW skymap alone. This method (hereafter \stratprob), which completely ignores galaxy catalogue information. Follow-up plans from electromagnetic telescopes often use the simple two-dimensional sky map $p_{\rm{2D}}$ for prioritising fields to be observed under practical constraints \citep[see for instance][]{Rana_2017}.

\subsection{Mass Based Pointing}\label{sec:strat_mass}

In the second strategy (\stratmass), we select all galaxies that are within the 99\% localisation contour of the GW event, and within a $\pm3 \sigma$ range of distances for that direction. We then prioritise the tiles by the galaxy mass contained in each of them, without convolving with the 2D or 3D probabilities. This is assuming that the heaviest galaxy is most likely to contain the GW source, irrespective of the GW localisation probability at that location. In this method, the GW probability is used as a binary cut for selecting galaxies --- beyond that, the location of the galaxy within the selected volume does not make any difference in its assigned weight. 

We then use the telescope tiling described in \S\ref{sec:simobs}, assigning each tile a priority based on the sum of all the galaxy weights inside it. Once the priorities are fixed, the observing scheme is to observe the top 100 tiles in this priority order.

\subsection{3D-probability $\times$ Mass based Pointing}\label{sec:strat_mp3}

The previous two sorting schemes ignored either the galaxy mass or the GW localisation probability at the galaxy location. In reality, both the metrics are important. Hence, in this scheme, we use set the weight of a galaxy to be $p_{\rm{3D}}\times m$, where $m$ is the mass of the galaxy, and $p_{\rm{3D}}$ is the GW probability at its location. This probability $p_{\rm{3D}}$ is obtained from the 3D distribution in the GW healpix maps, which depends on both the position and the galaxy distance. Again, we select all pixels within the 99\% localisation. We now evaluate a direction and distance dependent probability at the location of each galaxy based on distances provided by NED--LVS. The weight of each galaxy is calculated as the product of the mass and this 3D probability value. In Appendix~\ref{appendix:massprob}, we show that this normalised weight is indeed a probability density, and can be used directly in our calculations. Note that we do not have to apply an explicit cut for galaxy distances, as the 3D probability for galaxies at very low or very high distances from the median is vanishingly small.

This scheme (hereafter \stratmpthree) slightly improves on the number of galaxies that need to be targeted: about 8--10\% of the galaxies account for half the $p_{\rm{3D}} \times m$, as compared to 12--15\% required for just $m$ (\S\ref{subsec:ngal}, Figure~\ref{fig:top-heavieness}).

\begin{table*}[!th]
\centering
{\renewcommand{\arraystretch}{1.3}%
\begin{tabular}{|p{0.3\textwidth}|p{0.3\textwidth}|p{0.3\textwidth}|}  \hline
Follow-up Scheme & Filter & Weight \\ \hline
By raw 2D probability (\stratprob, \S\ref{sec:strat_prob}) & 99\% included area\newline No distance cut & 2D probability  \\  \hline
By mass only (\stratmass, \S\ref{sec:strat_mass}) & 99\% included area\newline Distance cut at median $\pm 3\sigma$ & Galaxy Mass \\  \hline
By 3D catalogue-based ranking (\stratmpthree, \S\ref{sec:strat_mp3}) & 99\% included area\newline No explicit distance cut & Galaxy mass $\times$ 3D probability \\   \hline
By 3D (filled mass + catalog) based ranking (\stratmassfill, \S\ref{sec:strat_massfill}) & 99\% included area\newline No explicit distance cut & (Galaxy mass + missing mass) $\times$ 3D probability \\
\hline
\end{tabular}
}
\caption{Summary of the five follow--up schemes discussed in this work. The ``Filter'' column describes any selection cuts applied in terms of sky localisation area and distances. The ``Weight'' column lists the quantity used for prioritising the galaxies.}
\label{table:comparisons}
\end{table*}

\subsection{Catalogue Incompleteness and Mass Filling}\label{sec:strat_massfill}

Galaxy catalogues have high completeness for nearby distances, but there is a lack of accurate redshifts for fainter galaxies at higher distances. \citet[Figure~10]{cook2023completeness} shows that comparing with a Near Infrared (NIR) luminosity function, NED-LVS is about 70\% complete up to a distance of 300--400~Mpc, beyond which the completeness fraction of galaxies with redshifts drops --- reaching 1\% at 1000~Mpc. This can become a major problem for any galaxy targeted approach for distant events. For the most massive and luminous galaxies ($L > L_*$), the catalogue is complete till $\sim 450$ Mpc. One can see that for a distant event (for instance at  800~Mpc), inherent biases of the catalogue would drive the observing decisions rather than the true locations of the galaxies.

We propose to overcome this limitation and account for the ``missing mass'' in the galaxy catalogue, by adding in the missing mass. From projected galaxy mass distribution and catalogue completeness as a a function of distance (from \citealt{cook2023completeness}), one can calculate the amount of mass that needs to be ``filled'' into any volume of space to compensate for this incompleteness. This can then be distributed uniformly in space, or we can create a mock catalogue of galaxies consistent with observations. \citet{2021JCAP...08..026F} have discussed some other possibilities as well. The mock catalogue approach is complex, and we would need a large number of mock catalogues to ensure a particular realisation of the mock catalogue does not bias our simulations. Furthermore, there is no objective way to decide which mock catalogue realisation should be used for final scheduling decisions. 

Instead, we proceed with the straightforward former approach: uniformly distributing the mass in the relevant volume of space. We divide space into concentric shells of 20~Mpc and calculate the total known galaxy mass in each shell from NED-LVS. Then, we scale this using fractional completeness data, given as a fraction for each distance shell (averaged over the sky) from \citet{cook2023completeness} to calculate the missing mass in each shell. 
For a given GW event, we divide the sky into 3D voxels, defined by these concentric shells in the radial direction and healpix pixels in the transverse direction, and calculate the mean missing (uncatalogued) mass of each 3D voxel.

If the voxel already contains a galaxy, then the mass of the voxel is set equal to the mass of the galaxy. Typically, $>$97\% voxels are empty --- for these, the mass is set to be the mean value of voxel mass for that 20~Mpc shell. Note that we assume the missing galaxy mass to be uniform all over the sky. A future work could address the case where we take the missing mass distribution to be direction--dependent. In effect, each voxel now contains either a galaxy from NED-LVS or appropriate artificially added mass to ensure catalogue completeness.

\begin{figure*}
    \centering
    \includegraphics[width = \linewidth]{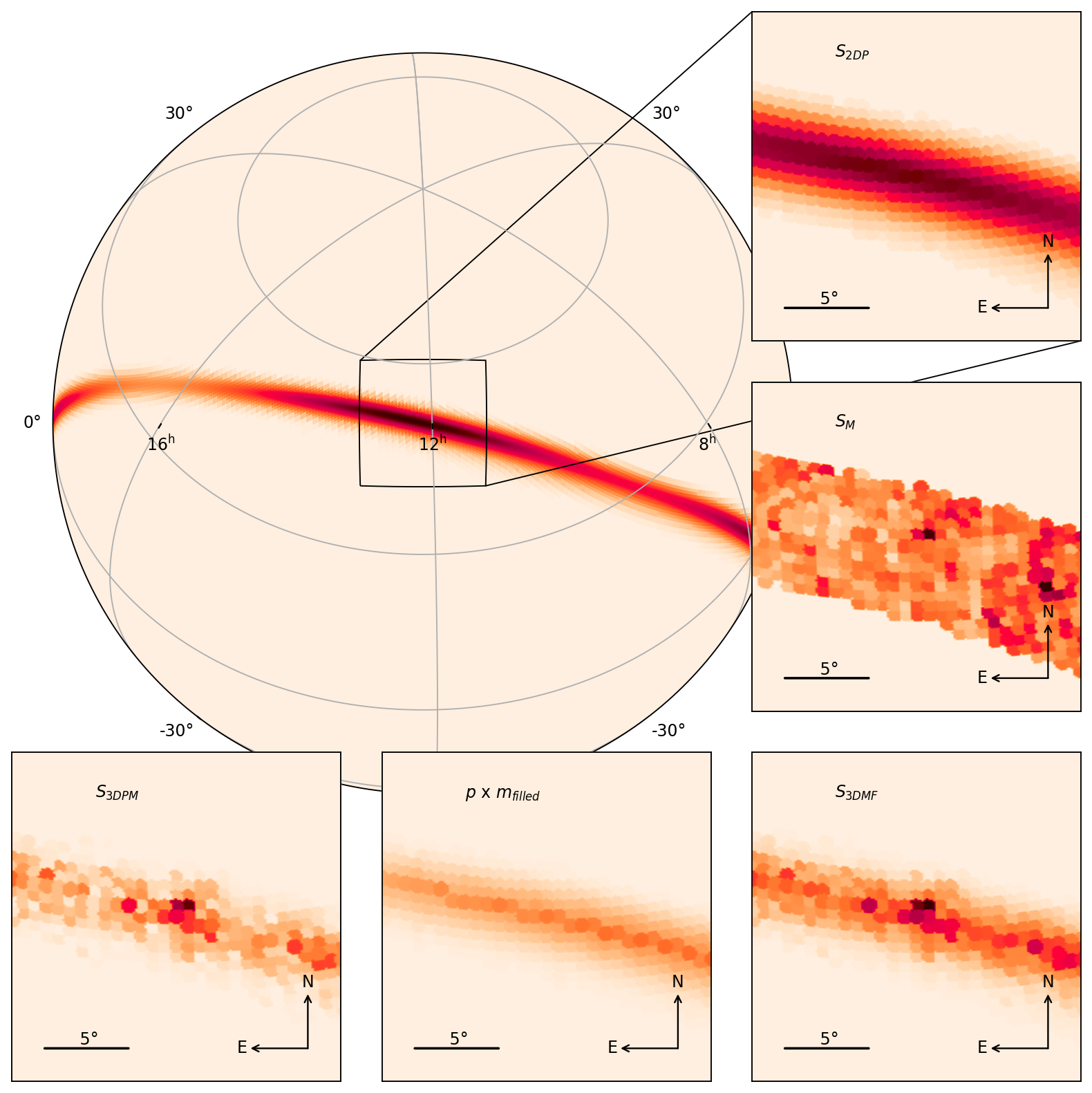}
    \caption{Skymap of a simulated event (number 1812 in the dataset). The globe shows the  probabilty density of the event localisation over the entire sky. The five insets show a zoomed-in plot for the central square.
    Circular markers in the top right inset denote the field of view of the GROWTH-India Telescope, with darker colours corresponding to a higher probability covered in that field: which becomes the basis of ordering observations in \stratprob. 
    The middle right inset shows the same GIT fields, but with the colour corresponding to the total catalogued galaxy mass as used in \stratmass.\\
    In the lower left inset, the fields are coloured as per the total probability weighted galaxy mass in that tile (\stratmpthree). The bottom centre inset uses the same colour bar range to show the probability-weighted missing mass \( p \times m_\mathrm{filled} \) in each tile. Adding this to the \stratmpthree\ map generates the final mass-filled weights (\stratmassfill) for each tile, which are denoted in the lower right plot with the same colour bar.\\
    From the insets, we can see that the ordering of observations is quite different for each strategy. Note that \( p \times m_\mathrm{filled} \) closely (but not exactly) follows the 2D probability of \stratprob. For this relatively close event (median distance 290~Mpc), the total contribution of missing mass is low, and the \stratmassfill\ map closely resembles the \stratmpthree\ map. If we repeat this exercise for more distant events, we find that the contribution from the missing mass starts increasing, and hence the \stratmassfill\ map starts resembling the \stratprob\ map.}
    
    \label{fig:skymap_globe_allfive}
\end{figure*}

We denote this final ``corrected mass'' of each voxel as $\hat{m}$. We then proceed with assigning weights following \S\ref{sec:strat_mp3}, replacing the $p_{\rm{3D}} \times m$ product with $p_{\rm{3D}} \times \hat{m}$ values. We refer to this scheme as \stratmassfill.

\section{Comparing with baseline observing strategies}\label{sec:results}

For each of the 329 events selected, we simulate observations with both telescopes using each of the five methods discussed in \S\ref{sec:strategies}. Figure~\ref{fig:skymap_globe_allfive} shows a GW sky map, with insets showing the weights for each GIT tile for each of the four methods. The globe in the centre shows the GW localisation sky map. The inset on the upper right shows a zoomed-in view of a square region of the globe. The circular markers each denote a field observed by GIT, and the colour denotes the total probability covered in that observation as per \stratprob. The middle right inset shows a simulated GIT tiling of the same sky map following \stratmass, with each circular patch denoting one tile, and darker tiles denoting higher masses. The bottom left inset shows tiling by \stratmpthree. The bottom centre inset is coloured as using the probability--weighted missing mass in each tile. The bottom right inset shows weights for tiling by \stratmassfill, and is obtained by adding the previous two. To help visual comparison, the three bottom insets use the same range for their colour bar. We see that each of the four schemes discussed can give quite different weights and hence priority orders for the tiles to be observed.

To quantitatively compare the performance of the strategies, we need to compare the result from each strategy with the ``true'' or ideal result. Among our five methods, the ``mass filled'' mock catalogue (\S\ref{sec:strat_massfill}) is the closest to such an ideal catalogue, and we assume that it represents the true mass distribution in the local universe. For instance, suppose we select 100 top tiles for GIT using \stratprob, and undertake a simulated observation. To calculate the actual probability coverage, we add up the $p_{\rm{3D}} \times \hat m$ for those hundred tiles, and use this number as the coverage for this event with the \stratprob\ method. A consequence of this choice is that \stratmassfill\ will always give the highest probability coverage in this work. 

Now, we consider \stratprob\ as the baseline method, and use simulated GIT observations to compare the other sorting methods with it in the following sub-sections.

\subsection{Comparing Galaxy Mass based Sorting (\stratmass)}\label{sec:comp_strat_mass}

\begin{figure*}
\centering
    \includegraphics[width = \linewidth]{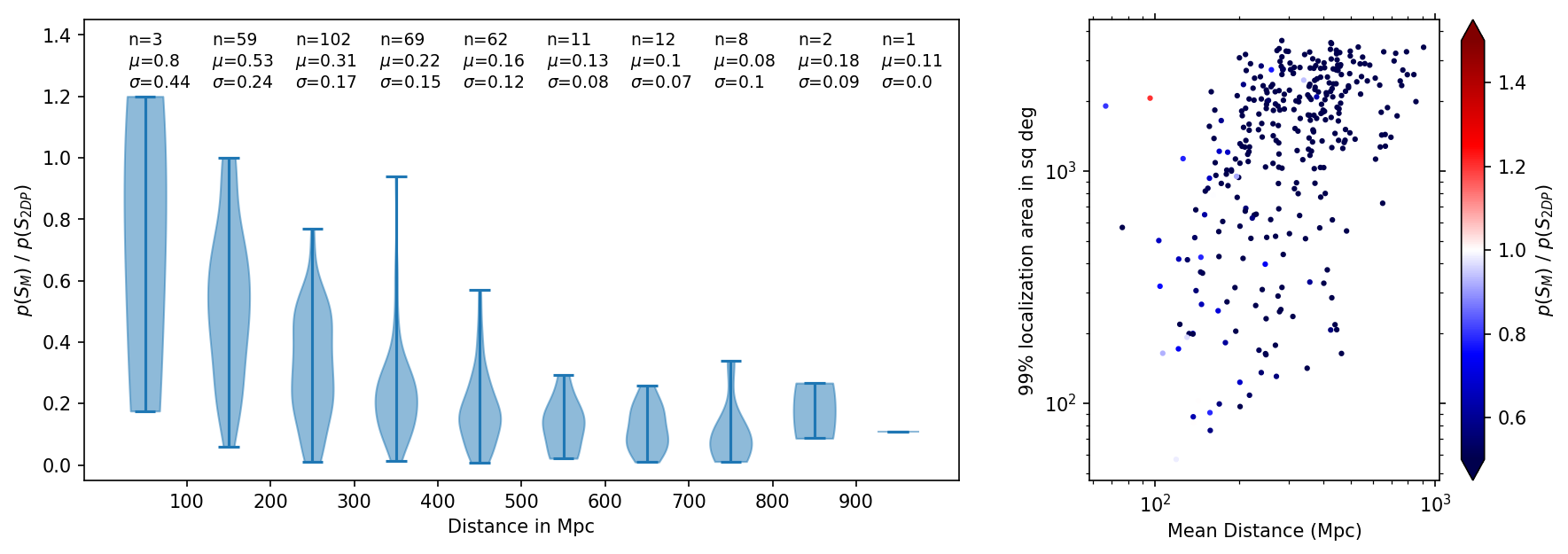}
    \caption{
    Left: Ratio of the probability-mass ($p\times \hat{m}$) covered using galaxy mass based ranking, to that covered with 2D probability-based ranking ($\mathcal{R} = p(\stratmass)/p(\stratprob)$), plotted for  100~Mpc distance bins (See \S\ref{sec:comp_strat_mass} for discussion). The three numbers at the top denote the number of events in that bin, and the median ($\mu$) and standard deviation ($\sigma$) of the ratios in that bin.
    Right: A scatter plot of the distance vs the $99\%$ localisation area for all the BNS and NSBH events studied, colour coded by the same ratio $\mathcal{R}$. We see that \stratprob\ performs significantly better ($\mathcal{R} \ll 1$) for most events, with the exception of a few nearby, poorly localised events. The colour bar is set to be the same for all figures of this form throughout this work, to allow for easy visual comparison. The colour bar is clipped to the range \(0.5 \le \mathcal{R} \le 1.5\), so dark blue symbols denote values \( \mathcal{R} \le 0.5\), while maroon denotes values \( \mathcal{R} \ge 1.5\). Here, the dark blue colour of most points highlights that \stratmass\ significantly under-performs as compared to \stratprob\ in most of the cases considered.}
    \label{fig:massonly}
\end{figure*}

First we look at the simplistic \stratmass\ (\S\ref{sec:strat_mass}) sorting. Simulations for observations with GIT confirm the expected answer: simply picking the most massive galaxies in the 99\% localisation area while ignoring the GW probability distribution results in very poor coverage of the total $p_{\rm{3D}} 
\times \hat m$ over the 99\% localisation area. We show these comparisons graphically in Figure~\ref{fig:massonly}. For each event, the distance used in the figure is the mean distance from the posteriors in its sky map. 
The top panel of Figure~\ref{fig:massonly} shows violin plots of the probability ratios $\mathcal{R} = p(\stratmass)/p(\stratprob)$ in 100~Mpc distance bins. We see that in most cases, the performance is a factor of few times lower than what can be achieved with \stratprob. The only exceptions are well-localised nearby events, where the localisation volumes only contain a handful of galaxies: hence all strategies end up observing almost all galaxies. The bottom panel shows \stratmass\ performs very poorly for events at large distances or large localisation areas.

Note that the method only considers galaxies within $3\sigma$ of the median distance in each pixel; without the distance cut the coverage obtained worsens.

\begin{figure*}[hp]
    \centering
    \includegraphics[width = \linewidth]{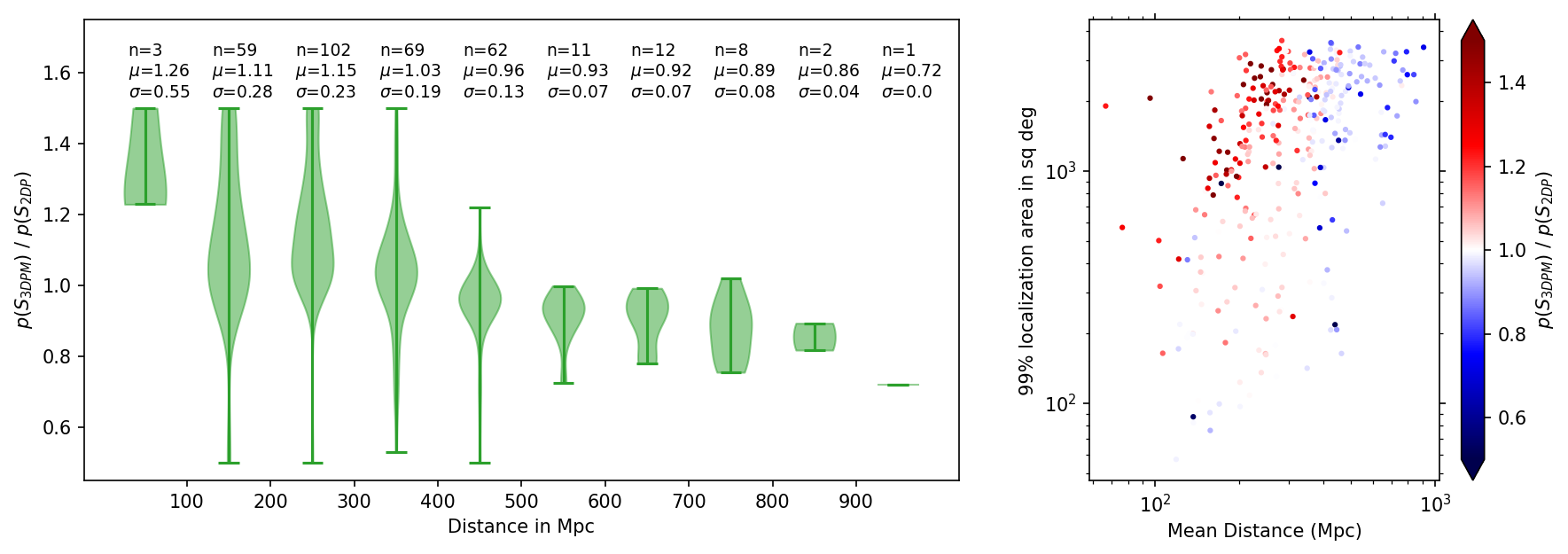} \\
    
    \caption{Similar to Figure~\ref{fig:massonly}, but comparing the 3D probability $\times$ mass strategy \stratmpthree\ with the baseline 2D probability tiling, using $\mathcal{R} = p(\stratmpthree)/p(\stratprob)$ (See \S\ref{sec:comp_strat_mp3} for discussion). We see that for distances $\lesssim 300-400$~Mpc, \stratmpthree\ performs better, while \stratprob\ yields better results ($\langle \mathcal{R} \rangle < 1$) at higher distances. The colour bar uses the same range as Figure~\ref{fig:massonly} and is clipped to \(0.5 \le \mathcal{R} \le 1.5\).
    }
    \label{fig:3dcat_vs_2draw}
\end{figure*}

\begin{figure*}
    \centering
    \includegraphics[width = \linewidth]{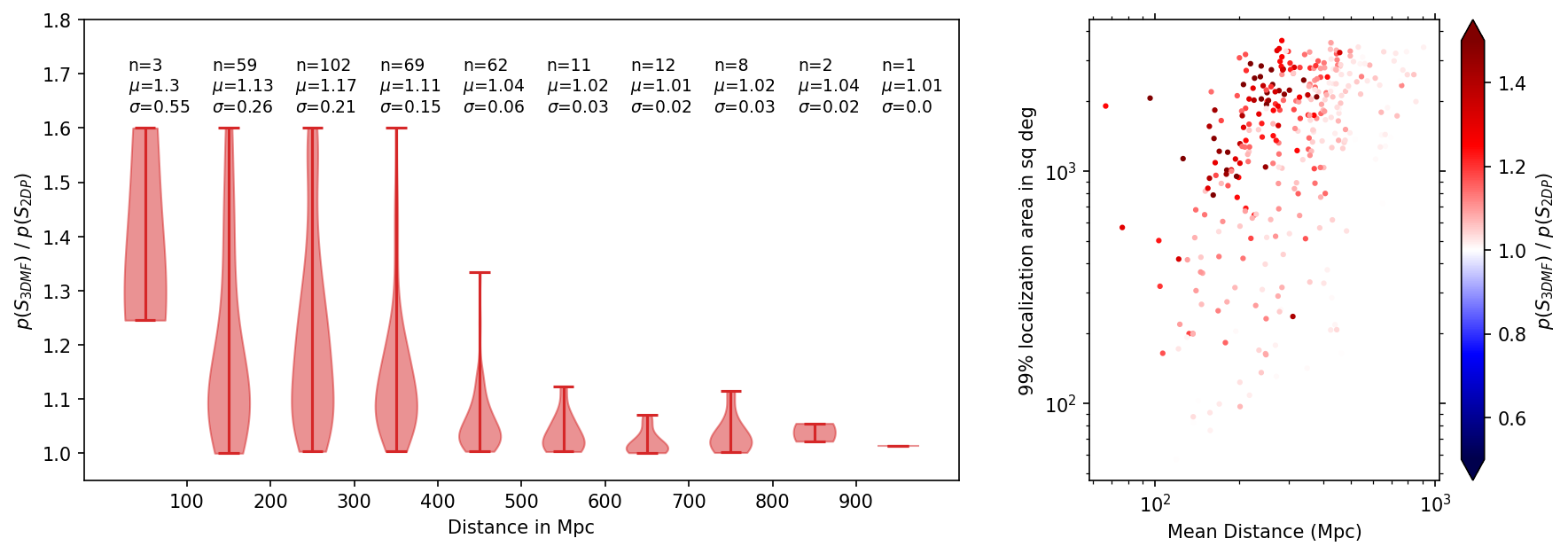}
    \caption{Similar to Figure~\ref{fig:massonly}, but comparing the 3D total probability $\times$ mass strategy \stratmassfill\ with the baseline 2D probability tiling, using $\mathcal{R} = p(\stratmassfill)/p(\stratprob)$ (See \S\ref{sec:comp_strat_massfill} for discussion). By definition, \stratmassfill\ performs better than \stratprob\ for all events, but they are nearly equal for distant events. The colour bar uses the same range as Figure~\ref{fig:massonly} and is clipped to \(0.5 \le \mathcal{R} \le 1.5\).}
    \label{fig:pmnet_vs_2draw}
\end{figure*}

\subsection{Comparing 3D-probability $\times$ Mass based Sorting (\stratmpthree)}\label{sec:comp_strat_mp3}
Next, we compare the strategy using galaxy catalogues and the 3D localisation probability (\stratmpthree). Here, we expect that \stratmpthree\ will perform better for nearby events where the galaxy catalogues are complete: ensuring that we target those earlier. However, at higher distances as the galaxy catalogues become increasingly more incomplete, using just a 2D sky map may give better results. Indeed, this is what the simulations show us. 
Figure~\ref{fig:3dcat_vs_2draw} shows the distribution of the ratio $\mathcal{R} = p(\stratmpthree)/p(\stratprob)$. 
We see that the bin-wise median value of the ratio is $>1$ for distances below 300~Mpc, but drops below 1 at higher distances.
The same trend is seen in the colours in the bottom panel of Figure~\ref{fig:3dcat_vs_2draw}, where red symbols (denoting better performance of \stratmpthree) dominate lower distances, while blue symbols denoting better performance of \stratprob\ dominate distances above $\sim 300 - 400$~Mpc. 

We also note that the advantage of using a galaxy catalogue is higher for larger localisation areas.

\subsection{Comparing Mass Filling based Sorting (\stratmassfill)}\label{sec:comp_strat_massfill}
Finally, we compare the results when using the actual mass-filled catalogue, accounting for the missing mass. We note that since we have used $p_{\rm{3D}} \times \hat{m}$ as the figure of merit, \stratmassfill\ will always perform better than other strategies as evaluated in this work --- the question is how strong the difference is. Figure~\ref{fig:pmnet_vs_2draw} shows that at distances below about 400--500~Mpc, there is a significant gain of up to $\mathcal{R} = 1.3$ by using the mass-filled catalogue. At higher distances, the methods become equivalent a majority of the mass is uncatalogued, and the ``missing mass'' part of the total weight traces the 2D probability.

This can also be understood by inspecting Figure~\ref{fig:skymap_globe_allfive}. The event shown in that figure (event ID 1812) has a median distance of 290~Mpc, and the galaxy catalogue is reasonably complete within the localisation volume. Due to this, the values of probability--weighted missing mass (bottom center inset) are often lower than those for known catalogued galaxies (\stratmpthree, bottom left inset). As a result the final \stratmassfill\ map is similar to the \stratmpthree\ map. On the other hand for more distant events, the contribution of missing mass increases, and the \stratmassfill\ maps starts resembling the \stratprob\ map.

\begin{figure*} 
    \centering
    
    \includegraphics[width = \linewidth]{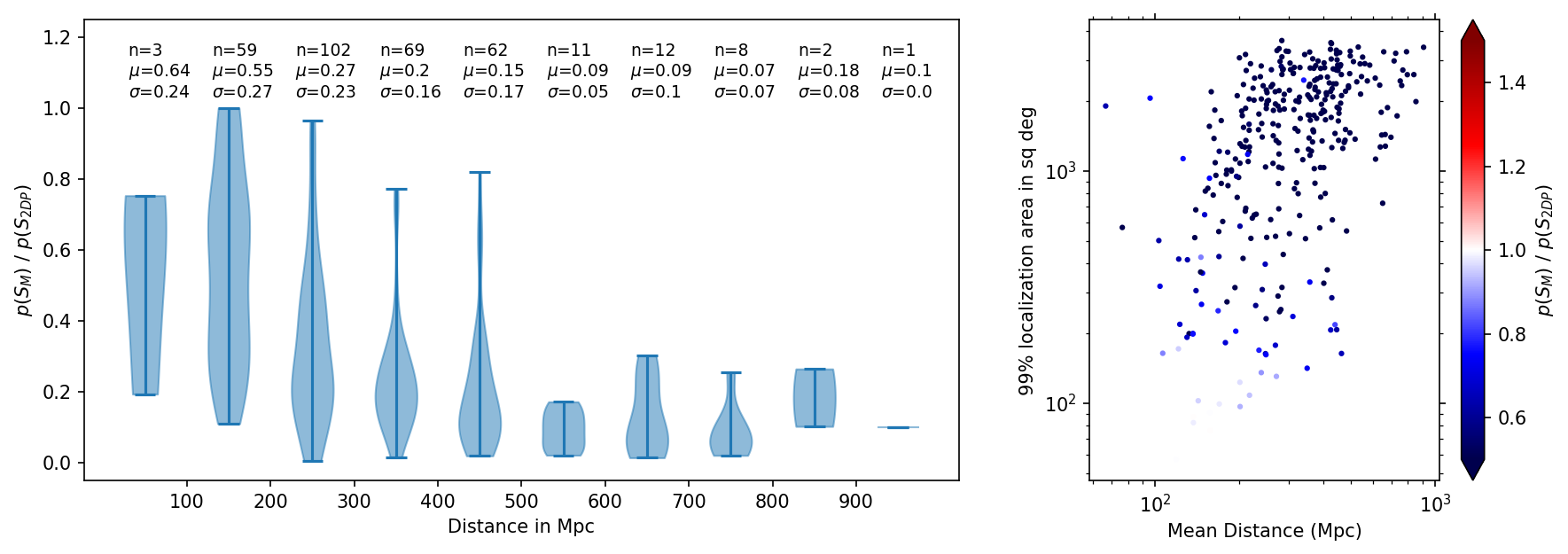} \\ 
    \includegraphics[width = \linewidth]{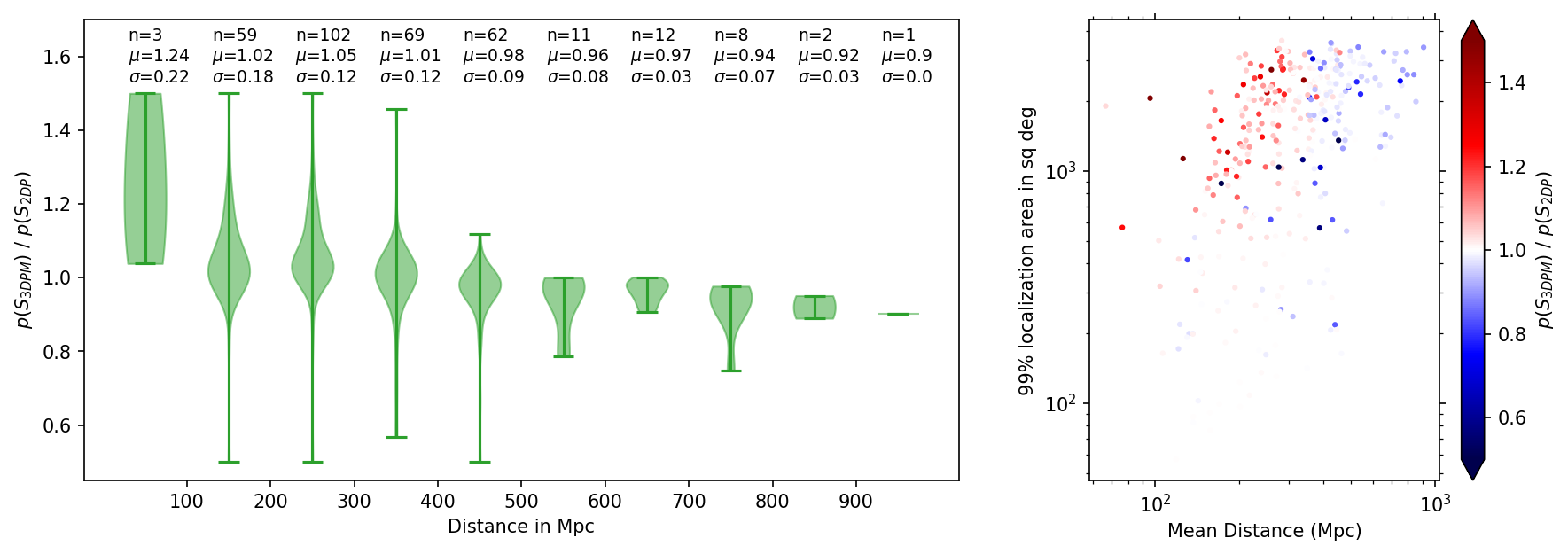} \\ 
    \includegraphics[width = \linewidth]{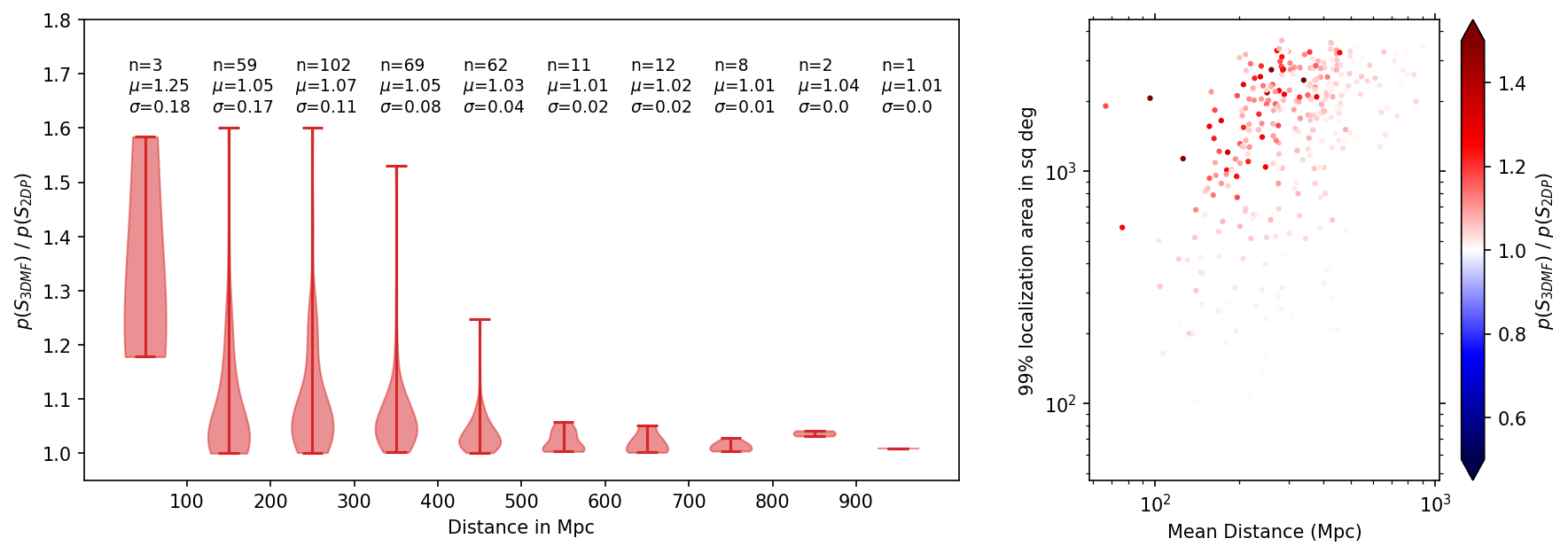} \\ 
    \caption{Same as Figures~\ref{fig:massonly},  ~\ref{fig:3dcat_vs_2draw} and ~\ref{fig:pmnet_vs_2draw}, but for WINTER tiling. The three left panels from top to bottom show violin plots for $\mathcal{R} = p(\stratmass)/p(\stratprob)$,  $p(\stratmpthree)/p(\stratprob)$, and $p(\stratmassfill)/p(\stratprob)$ respectively. The corresponding right panels show scatter plots between the localisation area and median distance, with the colour denoting $\mathcal{R}$. The colour bars use the same range as previous figures, and are clipped to \(0.5 \le \mathcal{R} \le 1.5\). As compared to GIT, we note that all ratios here are closer to unity, as the larger FoV leads to greater coverage. We also see that the changeover from better performance from \stratmpthree\ to \stratprob\ happens at the same distance range of 300--400~Mpc. Both these effects are as expected (\S\ref{sec:discuss}).}
    \label{fig:winter_coverage_ratio_plots}
\end{figure*}

\subsection{Impact of increased FoV}\label{sec:impact_fov}
We then repeat the exercise with WINTER, which has a 2.5 times higher FoV than GIT, and find similar trends (Figure~\ref{fig:winter_coverage_ratio_plots}). We see the same trends as GIT: galaxy catalogue based strategies perform better for near events where the catalogue is more complete, but \stratprob\ remains a better choice for distant events. 
The distance at which this changeover happens is not impacted by the larger FoV: this is consistent with our expectations that this distance depends only on catalogue completeness. We also find that the probability ratios (for instance $\mathcal{R} = p(\stratmpthree)/p(\stratprob)$ in the second panel, etc) are closer to unity for WINTER as compared to GIT. This too is expected: a telescope with a smaller field of will cover the highest probability regions. A wider field telescope will cover those regions and some more. This additional coverage of naturally contains lower probability regions, and gives diminishing returns.

\begin{table*}[!ht]
\centering
{\renewcommand{\arraystretch}{1.3}%
\begin{tabular}{|l|c|c|}  \hline
Follow-up Scheme & $d_\mathrm{GW} <$ 302~Mpc & $d_\mathrm{GW} >$ 302~Mpc \\ \hline
By raw 2D probability (\stratprob, \S\ref{sec:strat_prob}) & 0.458 (0.259, 0.733) & 0.316 (0.217, 0.487)  \\ 
By mass only (\stratmass, \S\ref{sec:strat_mass}) & 0.117 (0.066, 0.319) & 0.046 (0.029, 0.069) \\ 
By 3D catalogue-based ranking (\stratmpthree, \S\ref{sec:strat_mp3}) & 0.504 (0.326, 0.759) & 0.294 (0.215, 0.449) \\  
By 3D (filled mass + catalog) based ranking (\stratmassfill, \S\ref{sec:strat_massfill}) & 0.530 (0.340, 0.793) & 0.332 (0.243, 0.515) \\
\hline
\end{tabular}
}\caption{
Comparing typical coverage numbers for the GROWTH-India Telescope with the various follow-up strategies. The events are divided into two groups based on whether the nominal distance estimate from the GW signal ($d_\mathrm{GW}$) is less than or greater than 302~Mpc. The first number in each column gives the median value of the total $p\times \hat{m}$ for the simulated events. The numbers in the brackets are the 25th and 75th percentile respectively.}
\label{table:comparisons-git}
\end{table*}

\begin{table*}[!ht]
\centering
{\renewcommand{\arraystretch}{1.3}%
\begin{tabular}{|l|c|c|}  \hline
Follow-up Scheme & $d_\mathrm{GW} <$ 302~Mpc & $d_\mathrm{GW} >$ 302~Mpc \\ \hline
By raw 2D probability (\stratprob, \S\ref{sec:strat_prob}) & 0.664 (0.431, 0,892) & 0.491 (0.359, 0.660)  \\ 
By mass only (\stratmass, \S\ref{sec:strat_mass}) & 0.168 (0.097, 0.480) & 0.066 (0.039, 0.102) \\ 
By 3D catalogue-based ranking (\stratmpthree, \S\ref{sec:strat_mp3}) & 0.698 (0.471, 0.890) & 0.464 (0.349, 0.629) \\  
By 3D (filled mass + catalog) based ranking (\stratmassfill, \S\ref{sec:strat_massfill}) & 0.729 (0.505, 0.917) & 0.510 (0.376, 0.665) \\
\hline
\end{tabular}
}\caption{
Same as Table~\ref{table:comparisons-git} but for WINTER tiling  }
\label{table:comparisons-winter}
\end{table*}

\section{Discussion}\label{sec:discuss}

The overall performance of each method for simulated GIT and WINTER observations is summarised in Tables~\ref{table:comparisons-git} and \ref{table:comparisons-winter} respectively. The first number denotes the median $p_{\rm{3D}} \times \hat{m}$ probability coverage for simulated events, while the numbers in parentheses denote the 25th and 75th percentiles respectively. The median distance for our 329 event sample is 302~Mpc. Coincidentally, the completeness of the NED-LVS catalog is also nearly constant at closer distances, but starts falling off beyond this range. As discussed in \S\ref{sec:results}, this is also the range at which the efficacy of galaxy catalogue--based methods begins to fall off. Hence, in these tables we divide the sample into ``near'' events with $d_\mathrm{GW} < 302$~Mpc and ``distant'' events with $d_\mathrm{GW} > 302$~Mpc.

The tables show the same trends for both telescopes. \stratmpthree\ outperforms other methods for near events, but gives sub--par performance for distant events: where the baseline \stratprob\ typically gives a few percent higher coverage. \stratmassfill\ gives a few percent higher coverage as compared to other methods. We note that the differences between the baseline \stratprob\ and the often recommended \stratmpthree\ are comparable to the difference between the leading of those two methods and \stratmassfill: suggesting that accounting for the missing mass by spreading it out uniformly is a strategy worth considering.

When switching to a telescope with a larger field of view, all probabilities increase, but by less than the 2.5 ratio of the FoVs. This is indeed expected due to the diminishing returns in observing schemes that prioritise highest weight tiles first (without accounting for other observing constraints): the first tiles observed have the highest weights. As a cross-check, we also simulated ambitious 250-tile observations with GIT, covering the same sky area as the 100-tile WINTER observations --- and obtained very similar numbers. On the other hand, the overall coverage for all methods also scales as expected if we limit GIT to 50 observations instead of 100, and the relative trends discussed above continue to hold true.

In conclusion, we find accounting for the catalogue incompleteness by spreading out the ``missing mass'' gives non--trivial gains in the net probability coverage. Among the conventional methods, we recommend using the complete 3D probability information from IGWN in combination with the NED-LVS galaxy catalogue for events closer than about 300--400~Mpc, while using the simple 2D sky map yields better results for more distant events. This distance cut stems from the completeness of the galaxy catalogues, and will get revised as catalogues improve. An advantage of this \stratmassfill\ is that it naturally transitions from being similar to \stratmpthree\ at low distances, to becoming close to \stratprob\ at larger distances, where most of the mass is missing.

\subsection{Where are galaxy catalogues relevant?}
Further, we find that the gains obtained by using galaxy catalogues combined with the GW skymap are no longer as significant as they were in the earlier GW observing runs primarily due to there being a large number of galaxies in localisation volumes and the catalogue being incomplete at larger distances. Until deeper galaxy catalogues are created, at these large distances it suffices to tile the GW sky map using \stratprob\ to maximize the odds of discovering a counterpart. Even with deeper galaxy catalogues, the gains in probability covered are limited.

However, in any such tiled follow-up of a large sky area, a large number of candidate transients are identified. Considerable telescope resources need to be devoted to further follow-up of these candidates to obtain photometry to measure temporal evolution, and potentially spectroscopy to classify the source. The follow-up requirements can quickly explode beyond the available resources. It is here that galaxy catalogues are extremely crucial: the counterparts are expected to be within galaxies in the localisation volume. Hence, the limited telescope resources can be prioritised based on proximity to galaxies, and the 3D probability density or the (mass-)weighted probability density for that galaxy. This underscores the importance of galaxy catalogues in the search for the elusive electromagnetic counterparts to gravitational wave sources, and the need for continually updating and expanding the catalogues in the future.

\bibliography{galaxy}

\appendix

\section*{APPENDIX A: Mathematical Formulation for Probability-Mass}
\label{appendix:massprob}

Consider a GW event was detected by a detector with strain data $\strain$. We have a catalogue of galaxies that are potential hosts of this GW event. The catalogue contains on sky positions ($\alpha, \delta$), redshift $z$, and stellar mass estimates $M_*$ for all galaxies. We wish to find the probability that a galaxy $G$ in this catalogue is the host of a GW event corresponding to the strain data $\strain$, i.e., $P(G|\strain)$. If the intrinsic parameters of the event that corresponds to the strain data are $I$, we have,
\begin{align}
    P(G|\strain)  &= \int P(G, I| \strain) dI \\
    P(G, I|\strain) &\propto P(\strain|G, I)P(G, I) \,.\\
\end{align}
The first equation is a straightforward marginalization over intrinsic parameters $I$. The second equation corresponds to the Bayes' theorem, where $P(\strain|G, I)$ is the likelihood of observing the strain, $\strain$ given the intrinsic parameters corresponding to the strain data and the parameters $(\alpha, \delta, z)$ which closely correspond\footnote{Here we are neglecting the difference between the sky position of the galaxy and the gravitational wave event and the apparent redshift of the galaxy and the GW event. In principle we could also account for the difference if required by marginalizing over the possible locations of the GW event.} to the location of the galaxy $G$. The prior probability $P(G, I)$ corresponds to the joint probability of galaxy $G$ hosting an event with intrinsic parameters $I$. Since the strain data itself does not directly depend upon the other parameters of the galaxy such as its $M_*$, we will separate out the dependence explicitly,
\begin{align}
    P(\strain|G, I) = P(\strain| \alpha, \delta, z, I) \,.
\end{align}
The prior probability $P(G, I)$ can be written as
\begin{align}
    P(G, I) = P(I|G)P(G)\,
\end{align}
where $P(I|G)$ is the probability to have an event with intrinsic parameters $I$ in a galaxy $G$ and $P(G)$ the prior probability that the galaxy $G$ hosts a GW event. This probability is a combination of probabilities that a GW event happens at a particular location $(\alpha,\delta,z)$ and in a galaxy of stellar mass $M_*$. If we assume that these probabilities are independent\footnote{We could potentially also generalize this assumption, if required.}, then we can separate out the prior probability that a galaxy $G$ hosts a GW event into its component probabilities $P(G)=P(\alpha,\delta,z) P(M_*)$, thus resulting in
\begin{align}
P(G|\strain)  \propto \int P(\strain| \alpha, \delta, z, I) P(I|G) P(\alpha, \delta, z) P(M_*) dI
\end{align}

If we assume $P(I|G)$ is independent of $G$\footnote{This implies that the type of the galaxy does not dictate the intrinsic parameters of the possible GW events.}, then we have,
\begin{align}
    P(G|\strain)  &\propto \int P(\strain| \alpha, \delta, z, I) P(I) P(\alpha, \delta, z) P(M_*) dI\,.\\
&=  P(M_*) \int P(\alpha, \delta, z| \strain, I) P(I) dI\,,\\
&= P(M_*) P(\alpha, \delta, z| \strain)
\end{align}
where we have identified the integral in the penultimate equation as the posterior localisation probability provided by the LIGO team marginalizing over the intrinsic parameters $I$ of the event, while $P(M_*)$ is the prior probability that a galaxy with stellar mass $M_*$ hosts a gravitational wave event.

In order to normalize this probability, we have to consider a complete set of galaxies which could potentially host the GW events. Some of them may be in the catalogue while others may be absent. The above probability when integrated over all such potential hosts should give unity. By specifying a functional form for $P(M_*)$, we should thus be able to normalize the above probability.

\end{document}